\documentclass[aps,pre,twocolumn,superscriptaddress]{revtex4}
\usepackage{graphicx,amsmath,amssymb}
\usepackage[usenames]{color}
\usepackage{indentfirst}
\usepackage{float}
\usepackage{mathrsfs}

\begin{document}

\title{Optimal building block of multipartite quantum battery}
\author{Tian-Ran Yang}
\affiliation{Department of Physics, Chongqing University, Chongqing 401330,
China}
\author{Yu-Yu Zhang}
\email{yuyuzh@cqu.edu.cn}
\affiliation{Department of Physics, Chongqing University, Chongqing 401330,
China}
\affiliation{Graduate School, China Academy of Engineering Physics,
Beijing 100193, China}
\author{Hui Dong}
\affiliation{Graduate School, China Academy of Engineering Physics,
Beijing 100193, China}
\author{Libin Fu}
\affiliation{Graduate School, China Academy of Engineering Physics,
Beijing 100193, China}
\author{Xiaoguang Wang}
\email{xgwang1208@zju.edu.cn}
\affiliation{Zhejiang Institute of Modern Physics, Department of Physics, Zhejiang
University, Hangzhou 310027, China}

\begin{abstract}
To take quantum advantage of collective effects in many-body system, we
design an elementary block for building multipartite quantum battery, which
enables charging an atomic ensemble with optimal numbers in a common thermal bath.
One achieves maximum free energy
as the stored energy in the steady state, which
is prior to each atom parallel charging independently.
It ascribes to quantum collective effects in the ensemble of atoms induced by
the competition between the coherent driving and decoherent dissipation. The corresponding
thermodynamic efficiency of the energy storage is analyzed.
The existence of the optimal elementary units of multipartite quantum battery provide a guideline for
designing a realizable charging scheme.
\end{abstract}

\date{\today }
\maketitle

\section{Introduction}
Recently, diligent efforts are devoted to explore the possibility of taking
advantage of quantum resources to achieve superior performances in the
energy conversion and storage with the control achievement on multipartite
quantum system~\cite{strasberg,watanabe,campisi}. Quantum battery (QB) is a
quantum system for storing energy supplied by an external source. The
battery exploits quantum effects for efficient charging in comparison to its
classical counterpart~\cite{Hovhannisyan2013,Friis2016,Friis17,farina19}. A
renewed effort is devoted to enhance charging of multipartite batteries in a closed system as a
consequence of quantum correlations in many-body systems,
which is known as collective effects~\cite{Ferraro2018,Campaioli2017,zhang2019,Binder2015}.

When a multipartite battery is subjected to a common thermal bath,
it gives rise to interesting phenomenons such as
an increasing entropy, which establishes a link to quantum thermodynamics~\cite{faist,skrzpczyk,bera,ma19}.
An emergence of collective effects in quantum thermodynamics like the free energy as the
stored energy and work extraction are attractive quantum phenomenon in open systems, while the influence of thermodynamics
on such quantum effects are overlooked. Many efforts have been devoted to investigate the QB for
the energy storage in the thermal environment using different charging protocols~\cite
{barra19,Farina18,pirmoradian19}. More recently, a
dissipative charging process of a battery was suggested ~\cite{barra19}, in
which an efficient thermodynamic equilibrium state is approached by work
extraction under cyclic unitary operations with help of auxiliary systems.
This engineering needs to find a unitary evolution for
the battery and the auxiliary system with a globally conserved quantity,
which is controlled by a post selected driving agent.

A harmonic driving as an external source
has been proposed as an alternative powerful charging field due to the tunable
driving frequency for maximal stored energy~\cite{zhang2019}. Inspired by
the advantage of the collective effects and the harmonic driving,
a multipartite QB in a common thermal bath,
collectively coupled to a harmonic driving field, is an attractive battery model
for optimal energy storage. In such many-body system,
the interplay between the coherent driving and decoherent charging
induced by the thermal dissipation plays an essential role in
the cooperative many-body effects for the charging. The question is that whether such collective
effects in many-body system be harnessed to improve thermodynamically meaningful features in the
driven-dissipative charging protocols.

In this paper, we present the irreversible thermodynamic charging
of an ensemble of $N$ two-level atoms charged by
a harmonic driving field in a common thermal bath. It is different
from previous entropy-preserving or energy-conserving charging process under
unitary operations. With the increment of
atoms, the quantum collective effects
lead to a  non-monotonic behavior of  free energy and a decreasing entropy per atom,
which are induced by the competition between the coherent
driving force and the decoherent dissipation of the common thermal bath.
We find the optimal number of atoms as an elementary unit of the QB interacting with a
common thermal bath, which results in
maximizing the stored energy per atom. Engineering such optimal-atom battery
as a building block one achieves more free energy by compared to parallel charging
for $N$ independent atoms. Meanwhile, thermodynamic efficiency of the
energy storage in terms of the work done by the external charging field is analyzed.
\begin{figure}[tbp]
\includegraphics[trim=35 200 30 80,scale=0.35]{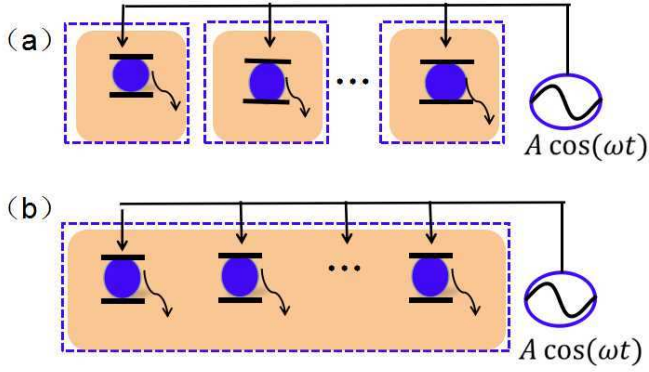}
\caption{(a)Charging protocol of $N$ two-level atoms in parallel. Each atom
is charged by a harmonic driving and couples to a thermal environment.
(b)The elementary building block (blue box) consists of a few atoms with a
common thermal bath. During the charging time, the QB
interacts with a harmonic driving field $A\cos(\protect\omega t)$.}
\label{charging}
\end{figure}


\section{A protocol for a multipartite QB charging }
We consider an open
charging system of a multipartite battery, which consists of two-level atoms
coupled to an external driving as a charger to transfer energy.
Fig.~\ref{charging} (a) shows normal parallel charging strategy with
independent thermal bath. Our charging protocol focus on the elementary
building block (blue box), illustratedin Fig.~\ref{charging} (b), with
finite number of atoms with shared thermal bath. Atoms in each units are
collectively charged by a harmonic field. The total Hamiltonian consists of
the QB-system part and the interacting part as $H=H_{s}+H_{I}$ with
\begin{equation}
H_{s}=\omega _{0}J_{z},H_{I}=A\cos (\omega t)J_{x},
\end{equation}
where $J_{\alpha }=\sum_{i}^{N}\sigma _{i}^{\alpha }/2$ ($\alpha =x,y,z$) is
the collective operator of $N$ two-level atoms with the energy level
splitting $\omega _{0}$. $A$ and $\omega $ are the driving amplitude and the
modulated frequency.

In the practical application, the QB are coupled also to the envirement,
which is modeled as the thermodynamics dissipation. The temporal
evolution of the density matrix $\rho(t) $ for the QB is described by the master
equation,
\begin{eqnarray}
\frac{d\rho (t)}{dt} &=&-i[\omega _{0}J_{z}+A\mathtt{cos}(\omega
t)J_{x},\rho ]  \notag \\
&&+\gamma \lbrack n(T)+1](2J_{-}\rho J_{+}-\{J_{+}J_{-},\rho \})  \notag \\
&&+\gamma n(T)(2J_{+}\rho J_{-}-\{J_{-}J_{+},\rho \}),
\end{eqnarray}%
where $n(T)=[e^{\omega /(k_{B}T)}-1]^{-1}$ is the mean occupation number of
bath mode with the frequency $\omega $ at the bath temperature $T$. The
Boltzmann's constant is set by $k_{B}=1$ for later discussion. The first
term on the right side describes the normal-parallel charging for $N$
independent atoms.
The basis set for representing the QB system of $N$ atoms
is the Dicke states $|J,l-N/2\rangle $ ($l=0,1,...,N$), which are
eigenstates of $J^{2}$ and $J_{z}$ with the total pseudo-spin $J=N/2$. Both
external driving and the dissipation induce the state transitions among
energy levels of the Dicke states $|N/2,l-N/2\rangle $ with $l=0,1,...N$.

\section{An optimal elementary unit for charging}
The importance is
to define the usable stored energy. In the normal
charging without thermal environment, the internal energy change of QB can
be utilized in the later retraction. Yet, only part of the internal energy
can be extracted in the charging process within the thermal environment. For such
evaluation, we have considered the entropy $S(\rho_{s})=-\mathtt{Tr}%
(\rho_{s}\ln\rho_{s})$ of the QB system. The useful energy stored in the QB
is measured by the free energy
\begin{equation}
F(\rho_{s})=E(\rho_{s})-k_{B}TS(\rho_{s}),
\end{equation}
where $E(\rho_{s})=\mathtt{Tr}(H_{s}\rho_{s})$ is the internal energy of the
QB system. The difference in free energy $\Delta F=\Delta E-k_{B}T\Delta S$
measures the useful energy stored in the QB. At zero temperature $T=0$, the
free energy change is equivalent to the mean energy, $\Delta
F(\rho_{s})=\Delta E(\rho_{s})$. Here we only consider the situation
with one thermal bath during the charging and later retraction process with the same temperature.
The similar definition of the useful work
is well discussed in the quantum  thermodynamic resource discussions \cite{alicki1979}.

Initially, $N$ atoms decouple with the charging field, and the QB
system is prepared in the Gibbs thermal state of $N$ atoms, $%
\rho_{0}=e^{-H_{s}/(k_{B}T)}/Z$ with the partition function $Z=\mathtt{Tr}%
[e^{-H_{s}/(k_{B}T)}]$. For the single-atom QB, the temporal mean energy $%
E(\rho_{s})=\omega_{0}\langle\sigma_{z}\rangle/2$ is obtained
analytically as (see the appendix)
\begin{eqnarray}\label{energy1}
E(\rho_{s})/\omega_{0} & = & -\frac{\gamma^{2}\chi}{\gamma^{2}%
\chi^{2}+A^{2}/2}\{1-\frac{e^{-3t\gamma\chi/2}}{2\chi\gamma^{2}}  \notag
\\
& & \{[(2\chi\gamma^{2}(1+\alpha\chi)+\alpha A^{2}]\cos(\Omega t)  \notag
\\
& & -\frac{\gamma}{2\Omega}\lbrack2\gamma^{2}\chi^{2}(1+\alpha\chi)+A^{2}(4+%
\alpha\chi)]\sin(\Omega t)\}\},  \notag \\
\end{eqnarray}
where the oscillation Rabi frequency is $\Omega=\sqrt{A^{2}-
\gamma^{2}\chi^{2}/4}$ with $\chi=[1+2n(T)]$. The mean energy $E$ of the
QB becomes larger as the driving amplitude $A$ increases or the dissipation
rate $\gamma$ decreases. At the zero temperature with $n(T)=0$, when the
driving strength $A$ becomes larger than the dissipative rate $\gamma$, $E$
tends to be zero in the steady state and the corresponding stored energy $%
\Delta E$ equals to $\omega_{0}/2$. (See Appendix A)

For the large value of $N$, we use the Hosltein-Primakoff transformation in
terms of auxiliary bosonic operators $b^{\dagger }$ and $b$: $%
J_{z}=b^{\dagger }b-N/2$, $J_{+}=b^{\dagger }\sqrt{N}$ and $J_{-}=b\sqrt{N}$.
It gives $d\rho (t)/dt=-i[\omega _{0}(b^{\dagger }b-N/2)+A\sqrt{N}\mathtt{cos%
}(\omega t)(b^{\dagger }+b),\rho ]+\gamma N[n(T)+1](2b\rho b^{\dagger
}-\{b^{\dagger }b,\rho \}+\gamma Nn(T)(2b^{\dagger }\rho b-\{bb^{\dagger
},\rho \})$. The driving strength is proportional to $A\sqrt{N}$, while the
decay rate is proportional to $\gamma N$. Due to the competition between the external
driving force and the thermal dissipation, one would expect the existence of the efficient elementary unit of the
QB with optimal number of atoms for maximizing the stored energy.

\begin{figure}[tbp]
\includegraphics[scale=0.4]{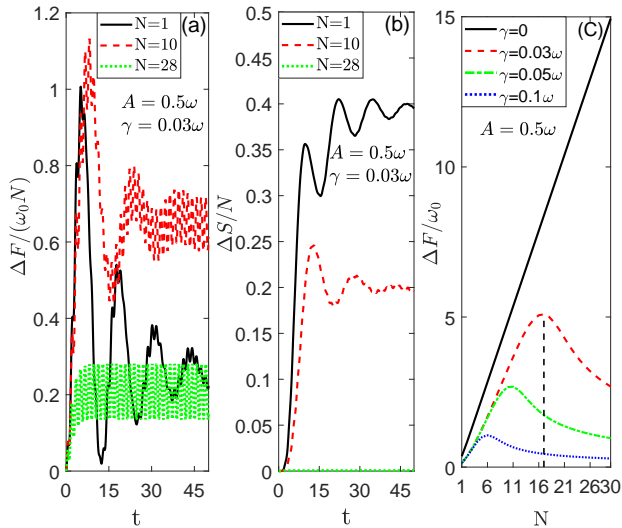}
\caption{(a)Variance of the scaled free energy $\Delta F/(N\protect\omega %
_{0})$ and (b) variance of the entropy $\Delta S$ as a function of the
charging time $t$ for different $N=1$, $10$ and $28$ with $A=0.5\protect%
\omega$ and $n(T)=0.2$. (c) $\Delta F/\protect\omega_{0}$ in the steady
state as a function of $N$ for different dissipation rate $\protect\gamma=0$%
, $0.03\protect\omega$, $0.05\protect\omega$ and $0.1\protect\omega$ with $%
A=0.5\protect\omega$.}
\label{Natom}
\end{figure}

Due to the difficulty of finding anylytical results for the multipartite batteries,
we present dynamiccal charging process via the free energy and the entropy by numerical calculations.
Fig.~\ref{Natom}(a) shows oscillations of the scaled free
energy change $\Delta F/N$, which exhibits a stable value at the steady state.
The charging period for the steady state becomes shorter as the atom number $N$ increases,
because the system relaxes to the steady state rapidly due to the dominated dissipation.
Interesting, the stored free energy in the steady state
increases firstly with the incremental of $N$, then decreases for $N=28$.
It demonstrates the existence of  the optimal number $N_{op}$
of atoms to achieve maximum free energy.  Meanwhile, it emergences inevitably an increasing of entropy
in the dissipative charging process, which is different from the preserved entropy in a closed system.
The corresponding
change of the scaled entropy $\Delta S/N$ decrease as $N$ increases in Fig.~\ref{Natom}(b).
It means that the heat flow induced by the entropy $Q_{s}=T\Delta S$
becomes lower than that obtained by $N$ atoms charging independently in parallel in Fig.~\ref{charging}(a).
It exhibits an improvement over the parallel charging of $N$ atoms.

One expects a steady state with larger population in high-level states to
achieve maximal stored energy. $N$ atoms with a strong driving amplitude $A$ would
result in the  occupation of  the higher-energy states $|N/2,l-N/2\rangle$ with a larger
value of $l$, yet the occupation of the lower-energy states with the increasing
of dissipation $\gamma$. As a consequence of such competition between the
coherent external driving and dissipation, an efficient elementary unit of
the QB with the optimal atoms is predicted to achieve maximum free energy
per atom. Fig.~\ref{Natom}(c) shows that the
non-monotonic dependence of the free energy $\Delta F$  on the number
of atoms $N$ in the steady state, exhibiting an maximum value at an optimal
number $N_{op}$ of atoms. Especially, in the absence of the dissipation $%
\gamma=0$, the free energy is proportional to $N$, $\Delta F\propto N$,
which is consistent with previous results~\cite{Ferraro2018,zhang2019}.
For different dissipation rate $\gamma$ and the driving strength $A$,
the efficient elementary unit of the QB consists of the optimal number $N_{op}$ of atoms
in a common thermal bath in Fig.~\ref{charging}(b), which has maximum
stored energy by comparing to results of
atoms charging independently in parallel.

\section{Charging efficiency}

In the charging process, the stored free energy
is an important quantity to evaluate the performance. However, we have shown the
energy from the agent dissipates into the thermal bath. The extent, to which the
work done by the agent is stored, is also important and typically evaluated
as the efficiency.  We quantify the efficiency of the charging process by the ratio
\begin{equation}
\eta =\frac{\Delta F}{W},
\end{equation}%
where the amount of the work done on the QB by the external driving field is
given by
\begin{equation}
W=\mathtt{Tr}(\int_{0}^{\tau }\frac{dH}{dt}\rho dt)=-A\omega \int_{0}^{\tau
}\sin (\omega t)\mathtt{Tr}(J_{x}\rho )dt.
\end{equation}

\begin{figure}[tbp]
\includegraphics[scale=0.43]{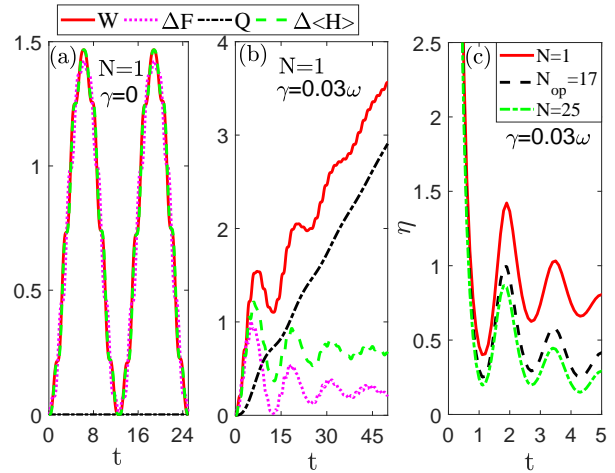}
\caption{Work done $W$, free energy $\Delta F$, variance of the internal
energy $\Delta\langle H\rangle$, and heat $Q$ as a function of time $t$ for $N=1$
quantum battery for different dissipation rate $\protect\gamma=0$ (a) and $\protect%
\gamma=0.03\protect\omega$ (b). (c) The efficiency $\eta$ as a function of time $t$
for different number of atoms $N=1$, $17$ and $25$ with $\gamma=0.03\omega$.
The parameters are $A=0.5\protect\omega$ and $n(T)=0.2$.}
\label{WQ}
\end{figure}
Due to the thermal dissipation of the irreversible process, some amount of
work $W$ is transferred into heat $Q$ flowing into the thermal bath, which
is measured by $Q=\mathtt{Tr}(\int_{0}^{\tau }\dot{\rho}Hdt)$.
The intrinsic energy of the charging system including the
interactions driving part $H_{I}$ is measured as $\langle H\rangle=\mathtt{Tr}%
[H(\rho(t)]=E+\langle H_{I}\rangle$, which is different from the mean energy $E$ of the QB
system.

Fig.~\ref{WQ} shows behaviors of the work $W$ done by the external driving force,
the heat flowing to the common bath $Q$.
For a closed system with $\gamma=0$ in Fig.~\ref{WQ}(a), the work $W$ is fully
transferred into the variance of the intrinsic energy $\Delta\langle
H\rangle $ periodically without heat dissipation.
One interesting finding is that the variance of the intrinsic energy $\Delta\langle H\rangle$
sometimes is lower than the variance of free energy $\Delta F=\Delta E$
due to the negative value of the interacting energy $\Delta \langle H_{I}\rangle<0$, resulting in the efficiency $\eta>1$.
Such efficiency is caused by the energy exchange between the coherent driving field and the QB.
For the dissipation case with $\gamma=0.03\omega$ in Fig.~\ref{WQ}(b), the laws of thermodynamics, $W=Q+\Delta H$,
holds in the open charging system as well as the closed system.
The corresponding efficiency $\eta$ decreases as the number $N$ of atoms
for an elementary unit of the QB increases in Fig.~\ref{WQ}(c). The efficiency for 
the efficient elementary unit with the optimal number of atoms $N_{op}=17$ is worse than that 
of single-battery case, which ascribes to heat dissipation.

\section{Conclusion}We have shown the driven-dissipative charging
of the multipartite battery comprising an ensemble of two-level atoms,
which interact with an external harmonic driving and a common
thermal bath. The stored energy in the battery is measured by the free energy
due to increasing of the entropy in
the open charging system. Due to the competition between the coherent driving
and the decoherent dissipation in atoms system, the free energy per atom in the steady state
behaves non-monotonically dependent on the number of the
atoms. We find the optimal elementary unit of the QB with finite atoms in a common thermal bath,
which provides maximal average stored energy.
To character the thermodynamic efficiency of the
charging process, we elucidate the work done by the external field as well as
the intrinsic energy of the QB system. We
prove the first law of thermodynamics holds in the open charging system.
The quantum thermodynamics in the open charging system involves
quantum collective effects induced by quantum coherence in many-body system
in comparison to classical thermodynamics, resulting in optimal building block of multipartite quantum battery.
Our results is a fundamental attributions to the powerful
energy storage with optimal atoms for physically realizable charging
schemes.

\acknowledgments
This work was supported by the NSFC
under Grants No.12075040, the Chongqing Research Program of Basic Research
and Frontier Technology under Grants No.cstc2020jcyj-msxmX0890, and
by the Fundamental Research Funds for the
Central Universities under Grant No. 2020CDJ-LHZZ-012.
H. Dong acknowledges the support from the NSFC (Grant No.11875049) and NSAF (Grant No.U1930403).
L. B. Fu acknowledges the support from the NSFC(Grant No.11725417) and NSAF (Grant
No.U1940403).
X. G. Wang acknowledges supports from
the NSFC (Grant No.11875231)
and the National Key Research and Development Program of China (Contracts No.
2017YFA0304202 and No. 2017YFA0205700).

\newpage
\appendix
\numberwithin{figure}{section}
\section{Analytical solutions for single-atom battery}
\begin{figure}[tbp]
\includegraphics[scale=0.4]{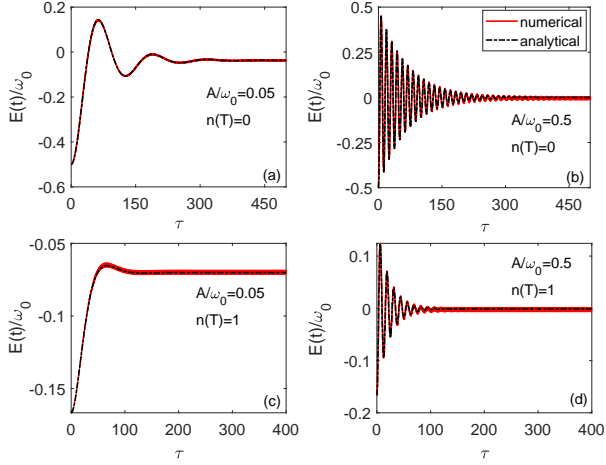}
\caption{Scaled mean energy $E/\protect\omega_0$ obtained by the analytical
solution (dashed black line) as a function of $\protect\tau$ for one-atom
quantum battery with different driving strength (a)(c) $A/\protect\omega%
_0=0.05$, and (b)(d) $A/\protect\omega_0=0.5$ at temperature $n(T)=0$ and $1$%
. The driving frequency is chosen as $\protect\omega=\protect\omega_0=2$,
and the dissipation rate is $\protect\gamma=0.01\protect\omega_0$. The
numerical results are listed for comparison (solid red line). }
\label{analytical}
\end{figure}

We analyze the thermodynamics of a single-atom battery charging process with
the master equation. The two-level system evolution involves energy transfer
from the external driving field and the dissipation of the energy into the
thermal environment.

The master equation of the single-atom battery system becomes
\begin{eqnarray}
d\rho _{s}/dt &=&-i\frac{\omega _{0}}{2}[\sigma _{z},\rho _{s}]-i\frac{A}{4}%
(e^{i\omega t}+e^{-i\omega t})[\sigma _{x},\rho _{s}]  \notag  \label{rho1}
\\
&&+\gamma \lbrack n(T)+1](2\sigma _{-}\rho _{s}\sigma _{+}-\{\sigma
_{+}\sigma _{-},\rho _{s}\})  \notag \\
&&+\gamma n(T)(2\sigma _{+}\rho _{s}\sigma _{-}-\{\sigma _{-}\sigma
_{+},\rho _{s}\}).
\end{eqnarray}

We perform a rotating-frame transformation using $U=\exp (i\omega t\sigma
_{z}/2)$ to give
\begin{eqnarray}
d\rho _{s}^{\prime }/dt &=&-i\frac{\omega _{0}-\omega }{2}[\sigma _{z},\rho
_{s}^{\prime }]  \notag \\
&&-i\frac{A}{4}(e^{i\omega t}+e^{-i\omega t})[e^{i\omega t}\sigma
_{+}+e^{-i\omega t}\sigma _{-},\rho _{s}^{\prime }]  \notag \\
&&+\gamma \lbrack n(T)+1](2\sigma _{-}\rho _{s}^{\prime }\sigma
_{+}-\{\sigma _{+}\sigma _{-},\rho _{s}^{\prime }\})  \notag \\
&&+\gamma n(T)(2\sigma _{+}\rho _{s}^{\prime }\sigma _{-}-\{\sigma
_{-}\sigma _{+},\rho _{s}^{\prime }\}),
\end{eqnarray}%
where $\rho _{s}^{\prime }=U\rho _{s}U^{\dagger }$. When the driving
strength $A$ is much smaller than the two-level energy $\omega _{0}$ on
resonance with the QB system $\omega _{0}=\omega $, it is reasonable to
making a rotating-wave approximation (RWA) by ignoring fast oscillating
terms. The Bloch equations are derived as
\begin{eqnarray}
\dot{\langle \sigma _{z}\rangle }_{\tau } &=&i\frac{A}{2}(\langle \sigma
_{-}\rangle _{\tau }-\langle \sigma _{+}\rangle _{\tau })-2\gamma \lbrack
\langle \sigma _{z}\rangle _{\tau }(2n(T)+1)+1],  \notag \\
&& \\
\dot{\langle \sigma _{+}\rangle }_{\tau } &=&i\frac{A}{4}\langle \sigma
_{z}\rangle _{\tau }-\gamma \langle \sigma _{-}\rangle _{\tau }(2n(T)+1), \\
\dot{\langle \sigma _{-}\rangle }_{\tau } &=&-i\frac{A}{4}\langle \sigma
_{z}\rangle _{\tau }-\gamma \langle \sigma _{+}\rangle _{\tau }(2n(T)+1).
\end{eqnarray}

For the initial Gibbs state with $\alpha =Tr[\sigma _{z}\rho (0)]$, it gives
analytically

\begin{eqnarray}
\langle \sigma _{z}\rangle _{\tau } &=&-\frac{2\gamma ^{2}\eta }{2\gamma
^{2}\eta ^{2}+A^{2}}\{1-e^{-3\tau \gamma \eta /2}\frac{1}{2\eta \gamma ^{2}}
\notag \\
&&\{[(2\eta \gamma ^{2}(1+\alpha \eta )+\alpha A^{2}]\cos (\Omega \tau )
\notag \\
&&-\frac{\gamma \lbrack 2\gamma ^{2}\eta ^{2}(1+\alpha \eta )+A^{2}(4+\alpha
\eta )]}{2\Omega }\sin (\Omega \tau )\}\},
\end{eqnarray}

where $\eta =[1+2n(T)]$, and the oscillation Rabi frequency is
\begin{equation}
\Omega =\sqrt{A^{2}-\frac{\gamma ^{2}\eta ^{2}}{4}}.  \label{frequency}
\end{equation}%
Especially, at zero temperature with $n(T)=0$ for $\langle \sigma
_{z}\rangle =-1$, $\langle \sigma _{-}\rangle =0$, $\langle \sigma
_{+}\rangle =0$, the corresponding temporal energy $E(\rho _{s})=\omega
_{0}\langle \sigma _{z}\rangle _{\tau }/2$ is obtained
\begin{eqnarray}
E(\rho _{s}) &=&\frac{-4\omega _{0}\gamma ^{2}}{8\gamma ^{2}+A^{2}}[1+\frac{%
A^{2}}{8\gamma ^{2}}e^{-3\gamma \tau /2}(\cos \Omega \tau +\frac{3\gamma }{%
2\Omega }\sin \Omega \tau )],  \notag  \label{energy0} \\
&&
\end{eqnarray}%
with $\Omega =\sqrt{A^{2}-\gamma ^{2}/4}$. When the driving strength $A$
becomes larger than the dissipative rate $\gamma $, $\langle \sigma
_{z}\rangle _{\infty }$ tends to be zero and the stored energy $\Delta E$
equals to $\omega _{0}/2$.

\begin{figure}[tbp]
\includegraphics[scale=0.45]{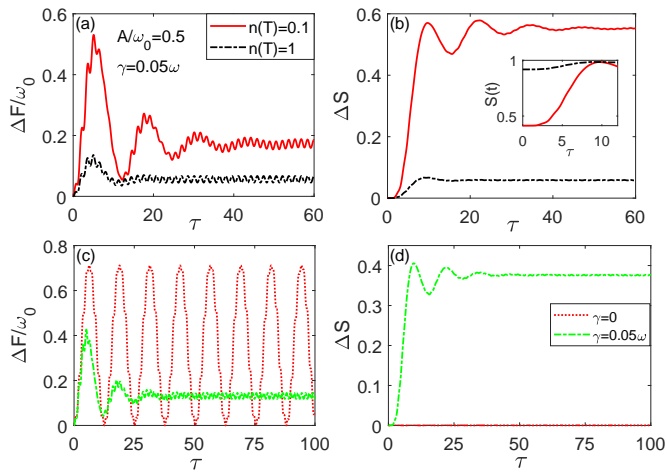}
\caption{(a)Scaled free energy $\Delta F/\protect\omega _{0}$ and (b) the Von Neumann entropy $
\Delta S$ of one-atom quantum battery for different temperature $n(T)=0.1, 1$ with dissipation rate $\gamma=0.05\omega$.
(c)Scaled free energy $\Delta F/\protect\omega _{0}$ and (d) the Von Neumann entropy
for different dissipative rate $\gamma=0$ and $0.05\omega$ with $n(T)=0.2$.
The driving frequency is chosen as $\protect\omega%
=\protect\omega_0=2$, and the driving strength $A/\protect\omega_0=0.5$.}
\label{rate}
\end{figure}

We calculate $E(\rho _{s})$ according to Eq.(\ref{rho1}) numerically without
the RWA for the dissipative charging process in Fig.~\ref{analytical}. The
analytical solutions in Eq.(\ref{energy0}) agree well with numerical ones
for $A/\omega _{0}=0.05$ and $0.5$. And the oscillation frequency of the
mean energy are correctly captured by the analytical one $\Omega $ dependent
on the driving strength $A$ and dissipative rate $\gamma $. As $A$ increases
to be much larger than $\gamma $, $A=0.5\omega _{0}$, $E(\rho _{s}) $
increases to be $0$ in the steady state in Fig.~\ref{analytical} (b), which
is consistent with the analytical ones from Eq.(\ref{energy0}). For a
thermal bath with $n(T)=1$, excellent agreement between the analytical
results and the numerical ones is observed in Fig.~\ref{analytical} (c) and
(d). As the temperature increases, the oscillation Rabi frequency $\Omega$
becomes smaller in Eq.(\ref{frequency}) due to the stronger thermal
dissipation. It demonstrates that the charging time to reach the steady
state becomes shorter at finite temperature, and the oscillation Rabi
frequency of the QB energy $E$ can be modulated by the driving strength $A$,
the loss rate $\gamma$ and the temperature $T$.

Fig.~\ref{rate} shows the thermodynamics charging controlled by the
temperature. Since the decay rate of the two-level atom is proportional to
the rate $\gamma \lbrack n(T)+1]$ and $\gamma n(T)$ in Eq.(\ref{rho1}). As $%
n(T)$ increases, the mean energy of the QB system and the free energy decay
rapidly as in Fig.~\ref{rate}(a). The corresponding variance of
the entropy $\Delta S$ decreases in Fig.~\ref{rate}(b). The increasing of the temperature induces
more energy flowing to the environment, while it reduces
the information flow. In the entropy non-preserving process, the heat flow
to the thermal bath decreases of the energy of the QB system due to
the entropy production. In the absence
of the dissipation $\gamma =0$, it exhibits periodic oscillation of the free
energy $\Delta F$ and the entropy is conserved in Fig.~\ref{rate}(c) (d).

\end{document}